\documentclass[%
 reprint,
 superscriptaddress,
 numerical,
 showpacs,
 amsmath,amssymb,
 aps,
 longbibliography,
 prd,
 floatfix
]{revtex4-1}
\usepackage{graphicx}

\usepackage[%
  colorlinks=true,
  urlcolor=blue,
  linkcolor=blue,
  citecolor=blue
]{hyperref}

\usepackage[pdftex]{color}
\usepackage[usenames,dvipsnames,svgnames,table]{xcolor}

\usepackage{mathdots}

\makeatletter
\DeclareFontFamily{OMX}{MnSymbolE}{}
\DeclareSymbolFont{MnLargeSymbols}{OMX}{MnSymbolE}{m}{n}
\SetSymbolFont{MnLargeSymbols}{bold}{OMX}{MnSymbolE}{b}{n}
\DeclareFontShape{OMX}{MnSymbolE}{m}{n}{
    <-6>  MnSymbolE5
   <6-7>  MnSymbolE6
   <7-8>  MnSymbolE7
   <8-9>  MnSymbolE8
   <9-10> MnSymbolE9
  <10-12> MnSymbolE10
  <12->   MnSymbolE12
}{}
\DeclareFontShape{OMX}{MnSymbolE}{b}{n}{
    <-6>  MnSymbolE-Bold5
   <6-7>  MnSymbolE-Bold6
   <7-8>  MnSymbolE-Bold7
   <8-9>  MnSymbolE-Bold8
   <9-10> MnSymbolE-Bold9
  <10-12> MnSymbolE-Bold10
  <12->   MnSymbolE-Bold12
}{}

\let\llangle\@undefined
\let\rrangle\@undefined
\DeclareMathDelimiter{\llangle}{\mathopen}%
                     {MnLargeSymbols}{'164}{MnLargeSymbols}{'164}
\DeclareMathDelimiter{\rrangle}{\mathclose}%
                     {MnLargeSymbols}{'171}{MnLargeSymbols}{'171}
\makeatother

\DeclareMathOperator{\diag}{diag}
\DeclareMathOperator{\tr}{tr}

\newcommand{\ket}[1]{| #1 \rangle}
\newcommand{\bra}[1]{\langle#1 |}

\begin{document}

\title{Supersymmetry in the nonsupersymmetric Sachdev-Ye-Kitaev model}

\author{Jan Behrends}
\affiliation{T.C.M. Group, Cavendish Laboratory, University of Cambridge, J.J. Thomson Avenue, Cambridge, CB3 0HE, United Kingdom}
\author{Benjamin B\'eri}
\affiliation{T.C.M. Group, Cavendish Laboratory, University of Cambridge, J.J. Thomson Avenue, Cambridge, CB3 0HE, United Kingdom}
\affiliation{DAMTP, University of Cambridge, Wilberforce Road, Cambridge, CB3 0WA, United Kingdom}

\begin{abstract}
Supersymmetry is a powerful concept in quantum many-body physics.
It helps to illuminate ground state properties of complex quantum systems and gives relations between correlation functions.
In this work, we show that the Sachdev-Ye-Kitaev model, in its simplest form of Majorana fermions with random four-body interactions, is supersymmetric. 
In contrast to existing explicitly supersymmetric extensions of the model, the supersymmetry we find requires no relations between couplings. 
The type of supersymmetry and the structure of the supercharges are entirely set by the number of interacting Majorana modes, and are thus fundamentally linked to the model's Altland-Zirnbauer classification.
The supersymmetry we uncover has a natural interpretation in terms of a one-dimensional topological phase supporting Sachdev-Ye-Kitaev boundary physics, and has consequences away from the ground state, including in $q$-body dynamical correlation functions. 
\end{abstract}

\maketitle

The Sachdev-Ye-Kitaev (SYK) model~\cite{Sachdev:1993hv,Kitaev2015} is a toy model that provides insight into diverse physical phenomena, ranging from the holographic principle~\cite{Maldacena:1999js,Gubser:1998id,Witten:1998ko} to quantum chaos~\cite{Shenker:2014ct,Maldacena:2016gp,Garcia:2017dv,Altland:2018gv,Iyoda:2018bv,Micklitz:2019kt}, and non-Fermi liquid behavior of strongly correlated electron systems~\cite{Davison:2017hf,Gu:2017hr,Song:2017hd,BenZion:2018fl,Altland:2019ec,Sachdev:2015dp,Patel:2018kf}.
Similar to black holes, it is believed to scramble quantum information with maximal efficiency~\cite{Bekenstein:1973ko,Sachdev:2015dp}.

The simplest variant of the SYK model describes $k$ Majorana fermions that interact through a random four-body term~\cite{Kitaev2015}. 
Its proposed physical realizations include mesoscopic systems based on Majorana fermions in vortices or quantum dots~\cite{Pikulin:2017js,Chew:2017fn}, or the ends of a one-dimensional topological phase~\cite{You:2017jj}.

Various generalizations of the SYK model have been considered, including models with $n$-body interactions~\cite{Maldacena:2016hu,Gross:2017fn} and supersymmetric extensions~\cite{Fu:2017hl,Li:2017bi,Kanazawa:2017be,HunterJones:2018hu,Garcia:2018jz}.
Typically, exact supersymmetry (SUSY) requires fine-tuning of the parameters~\cite{Fendley:2003ef,Fendley:2003ch,Hsieh:2016gj,Sannomiya:2017ib}.
In the supersymmetric SYK extensions, this fine-tuning corresponds to requiring certain relations between different couplings~\cite{Fu:2017hl}.

In this work, we show that already the simplest four-body SYK model, without any fine-tuning, is supersymmetric for all but two values of $k \mod 8$.
The type of SUSY depends only on $k$.
The supercharges will be shown to relate to ramps and long-time plateaus in time-dependent correlation functions~\cite{Cotler:2017fx}, which thus provide signatures of SUSY far from equilibrium.
In particular, we find that the number of supercharges is linked to the presence and nature of time-reversal symmetry and is reflected in the ramp shape~\cite{Guhr:1998bg}.
We also show that the number and structure of supercharges set the plateau features in $q$-body time-dependent correlation functions. 

Throughout this work, we focus on SUSY in the sense of supersymmetric quantum mechanics~\cite{Witten:1982cs,Fendley:2003ef,Fendley:2003ch,Nicolai:1976gf,Sukumar:1985bu,Forster:1989kf,Cooper:1995dp,Junker:1996ej,Correa:2007dq}.
SUSY is characterized by $\mathcal{N}$, the number of mutually anticommuting Hermitian fermionic supercharges that square to the Hamiltonian~\cite{Witten:1982cs}
\begin{align}\label{eq:SUSYdef}
 \lbrace Q_a , Q_b \rbrace = 2 H \delta_{ab} , & & [H,Q_a ] = 0 .
\end{align}

The Hamiltonian we consider describes four-body interactions between $k$ Majorana modes~\cite{Kitaev2015}
\begin{equation}
 H = \sum_{t =0}^{k-1} \sum_{s=0}^{t-1} \sum_{r=0}^{s-1} \sum_{q=0}^{r-1} J_{qrst} \gamma_q \gamma_r \gamma_s \gamma_t + E_0
 \label{eq:syk_hamiltonian}
\end{equation}
with real (as required by Hermiticity) but otherwise structureless couplings $J_{qrst}$, and the constant $E_0$ that ensures positive energies.
The Hermitian Majorana operators $\gamma_q = \gamma_q^\dagger$ satisfy the anticommutation relation $\lbrace \gamma_q , \gamma_r \rbrace = 2 \delta_{qr}$~\cite{Kitaev:2000gb}, and span an $M$-dimensional Hilbert space with $M=2^{\lceil k/2 \rceil}$~\footnote{Each pair of Majoranas spans a two-dimensional Hilbert space; for an odd number of Majoranas, we insert an additional Majorana $\gamma_\infty$ to obtain a fermionic Hilbert space, giving $M$ as above.}.
Since each term in the Hamiltonian~\eqref{eq:syk_hamiltonian} contains an even number of Majoranas, it conserves fermion parity $P$, given by
\begin{equation}
 P = \begin{cases}
 i^{k/2} \gamma_1 \gamma_2 \ldots \gamma_k & \text{even $k$} \\
 i^{(k+1)/2} \gamma_1 \gamma_2 \ldots \gamma_k \gamma_\infty & \text{odd $k$.}
\end{cases}
\end{equation}
To work in a Hilbert space with well-defined fermion parity, the additional Majorana $\gamma_\infty$ ``at infinity"
must be included when $k$ is odd~\cite{Fidkowski:2011dh}.
The operator $\gamma_\infty$ is not local to the SYK model; considering, e.g., a realization in a superconducting vortex~\cite{Pikulin:2017js}, it represents a degree of freedom with support far away from the vortex where the SYK Majoranas $\gamma_{j\neq \infty}$ reside. 
Like the local Majoranas, $\gamma_\infty$ is Hermitian and satisfies $\{ \gamma_q ,\gamma_r \} = 2 \delta_{qr}$.
Since $[H,P]=0$, all eigenstates of $H$ can by labeled by their parity eigenvalue $p=\pm 1$, giving $H \ket{\psi_\mu^p} = \varepsilon_\mu^p \ket{\psi_\mu^p}$ and $P \ket{\psi_\mu^p} = p \ket{\psi_\mu^p}$.

The number of interacting Majorana modes, specifically $k\mod 8$, sets the model's antiunitary symmetries~\cite{You:2017jj,Fidkowski:2011dh,Behrends:2019jc,Garcia:2016il}.
These come in two variants $T_\pm$, antiunitary operators satisfying $T_\pm \gamma_{q\neq \infty} T_\pm^{-1} = \gamma_{q\neq \infty}$.
They further satisfy $T_\pm P T_\pm^{-1}= \pm P$.
We refer to $T_+$ as time-reversal symmetry because it commutes with fermion parity and hence sets correlations within a parity sector.
Conversely, we call $T_-$ particle-hole symmetry.
Crucially for this work, since $T_-$ flips fermion parity, its presence implies correlations \emph{between} parity sectors.

The consideration of both $T_+$ and $T_-$ implies~\cite{Behrends:2019jc} a classification with more structure than the threefold Wigner-Dyson way highlighted in Ref.~\onlinecite{You:2017jj}.
In fact, as we now briefly review, it gives rise to the eight real Altland-Zirnbauer classes.
For even $k$, either time-reversal symmetry $T_+$ or particle-hole symmetry $T_-$ is present.
For odd $k$, both $T_+$ and $T_-$ are present; in this case $T_+ \gamma_{\infty} T_+^{-1} = (-1)^{(k+1)/2}\gamma_{\infty}$.
Their product, the unitary operator $Z= T_+T_-$, equals the product of all local Majorana operators up to a complex phase and corresponds to a chiral symmetry~\cite{Fidkowski:2011dh,Behrends:2019jc}.
A key feature of $Z$, which we will use repeatedly for diagnosing locality, is $[Z,\gamma_{q\neq\infty}]=0$ and $\{Z,\gamma_\infty\}=0$.
The squares $T_\pm^2$ vary with $k$ and label the eight real Altland-Zirnbauer classes~\cite{Altland:1997cg}.
While the Altland-Zirnbauer and Wigner-Dyson picture give the same level spacing statistics, the former also takes cross-parity correlations into account.
We summarize the symmetry classification in Table~\ref{tab:altland_zirnbauer} and review it in detail in Appendix~\ref{sec:eightfold}.

\begin{table}
 \begin{tabular}{c|cccccccc}
 \toprule
  $k \mod 8 $ & 0 & 1 & 2 & 3 & 4 & 5 & 6 & 7 \\
  \colrule
  Label & AI & BDI & D & DIII & AII & CII & C & CI \\
  \colrule
  $T_+^2$	& $+1$	& $+1$	& $0$	& $-1$	& $-1$	& $-1$	& $0$	& $+1$	\\
  $T_-^2$	& $0$	& $+1$	& $+1$	& $+1$	& $0$	& $-1$	& $-1$	& $-1$	\\
  \botrule
 \end{tabular}
 \caption{Time-reversal symmetry $T_+$ and particle-hole symmetry $T_-$ in the SYK model.
 The symmetries may be absent (denoted by $0$), or present and square to $-1$ or $+1$. 
 }
 \label{tab:altland_zirnbauer}
\end{table}

SUSY is known to imply a degeneracy between the parity sectors~\cite{Witten:1982cs}: the supercharges $Q_a$ exchange bosonic states with parity eigenvalues $p=+1$ and fermionic states with parity eigenvalue $p=-1$~\cite{Witten:1982cs}. Thus, the supercharges anticommute with fermion parity, $\{ P, Q_a \} =0$.
The presence of particle-hole symmetry also guarantees degeneracy between parity sectors, which as we now note, also implies SUSY. 
Parity degeneracy directly follows from particle-hole symmetry because $\ket{\psi^p_\mu}$ and $T_- \ket{\psi^p_\mu}$ have the same energy $\varepsilon_\mu=\varepsilon_\mu^p= \varepsilon_\mu^{-p}$ (since $[T_-,H]=0$), but opposite parity ($\lbrace T_-,P \rbrace = 0$)~\cite{Fidkowski:2011dh,Behrends:2019jc}.
Therefore, $\ket{\psi_\mu^p} \bra{\psi_\mu^{-p}}$ is an odd-parity zero mode, i.e., an operator that commutes with the Hamiltonian, but anticommutes with fermion parity~\cite{Behrends:2019jc}.
This in turn implies SUSY: The operator $\tilde{Q}_\mu = \sqrt{\varepsilon_\mu} \ket{\psi_\mu^+} \bra{\psi_\mu^{-}}$ satisfies $\tilde{Q}_\mu \tilde{Q}_\mu^\dagger = \varepsilon_\mu \ket{\psi_\mu^+} \bra{\psi_\mu^+}$ and $\tilde{Q}_\mu^\dagger \tilde{Q}_\mu = \varepsilon_\mu \ket{\psi_\mu^{-}} \bra{\psi_\mu^{-}}$, and hence the linear combinations $Q_{1,\mu} = \tilde{Q}_\mu + \tilde{Q}_\mu^\dagger$ and $Q_{2,\mu} = i (\tilde{Q}_\mu - \tilde{Q}_\mu^\dagger)$ are Hermitian, anticommute, and square to $\varepsilon_\mu$ times the projector on the two parity-degenerate states.
Consequently, the two supercharges
\begin{align}
 Q_1 = \sum_\mu (\tilde{Q}_\mu + \tilde{Q}_\mu^\dagger) , & & Q_2 = -i \sum_\mu (\tilde{Q}_\mu - \tilde{Q}_\mu^\dagger)
 \label{eq:supercharge_structure}
\end{align}
satisfy Eq.~\eqref{eq:SUSYdef} and anticommute with $P$.
Particle-hole symmetry is present unless $k=4n$. Thus, all but two of the symmetry classes are supersymmetric. 

Given the presence of six supersymmetric classes, there are a number of questions regarding the interplay of SUSY and the symmetry classification. How does $\mathcal{N}$ depend on the symmetry class? How do $Q_j$ transform under $T_\pm$ and how does this translate to the structure of the supercharges? We next turn to these questions.

We start with counting $\mathcal{N}$. A direct approach is based on counting level degeneracies. This follows from the observation that the ``spectrally flattened'' Hermitian supercharges $\Gamma_{j}=Q_{j}/\sqrt{H}$ satisfy
\begin{align}
 \{ \Gamma_{j},\Gamma_{k}\} = 2\delta_{jk}, & & [H,\Gamma_{k}] = 0 , & & \{P,\Gamma_{k}\}=0.
 \label{eq:Gamma}
\end{align}
They are thus many-body zero mode forms of Majorana fermions.
An even $\mathcal{N}$ of such zero modes give rise to a $2^{\mathcal{N}/2}$-dimensional fermionic degeneracy space for each of the $\varepsilon_\mu$ with one of the $|\psi_\mu^p\rangle$ chosen as ``vacuum''.
(With a suitable choice, the $\Gamma_j$-fermion parity of an eigenstate matches the state's physical fermion parity.)
This procedure is similar in spirit to the standard construction of supermultiplets~\cite{wess1992supersymmetry}.
For the six supersymmetric SYK classes, a twofold degeneracy is guaranteed by $T_-$ and a further twofold (Kramers) degeneracy is present whenever $T_+^2=-1$, resulting in an overall fourfold degeneracy.
This suggests $\mathcal{N}=2$, except for DIII and CII where this count gives $\mathcal{N}=4$.
What this counting does not address is how many $\Gamma_j$ (and hence $Q_j$) are local to the SYK model.
Next we investigate this to obtain the decomposition $\mathcal{N}=\mathcal{N}_\text{loc}+\mathcal{N}_\infty$ with $\mathcal{N}_\text{loc}$ counting the number of supercharges involving only $\gamma_{q\neq\infty}$. 
We first discuss the symmetry classes D and C before demonstrating the implications of locality in classes BDI and CI.
For brevity, we derive the supercharges in classes DIII and CII with $T_-^2 = -1$ in Appendix~\ref{sec:supercharges} and only summarize the results here.

We begin with classes D and C. Here $k$ is even hence all $\gamma_q$ are local. Therefore, our argument above applies directly: we find $\mathcal{N}=\mathcal{N}_\text{loc}=2$. 
All the other supersymmetric classes have odd $k$, thus potentially $\mathcal{N}\neq\mathcal{N}_\text{loc}$ due to $\gamma_\infty$. As we shall see, in all of these classes $\mathcal{N}=\mathcal{N}_\text{loc}+1$ with $\mathcal{N}$ following its degeneracy-based value above.
This is intuitive because $\gamma_\infty\equiv \Gamma_\infty$ automatically satisfies Eq.~\eqref{eq:Gamma} (in particular, it anticommutes with any local parity-odd operator), thus $\mathcal{N}_\text{loc}$ is at most $\mathcal{N}-1$.
To formally establish $\mathcal{N}_\text{loc}$, and the transformation of $\Gamma_{j}$ under $T_\pm$, we work in the energy eigenbasis, $H = \diag ( \lbrace \varepsilon_\mu \rbrace ) \otimes \openone_{2^{\mathcal{N}/2}}$, with $P = \openone_{M/2}\otimes\tau_3$.
(Here and below, $\tau_j$ and $\sigma_j$ are Pauli matrices; $\tau_j$ act in parity grading and $\sigma_j$ in the space of Kramers doublets, where applicable. We will often omit trivial tensor factors.)
In this basis, class D (C) has particle-hole symmetry [up to a phase $\diag ( \lbrace \exp(i\varphi_\mu) \rbrace )$ omitted here and below] $T_- = \tau_{1(2)} \mathcal{K}$ (with $\mathcal{K}$ for complex conjugation);
this follows from $T_-^2=\pm1$ and parity being the only degeneracy, $\openone_{2^{\mathcal{N}/2}}=\tau_0$.
We have $\Gamma_{1,2}=\tau_{1,2}$, which correspond to the two supercharges introduced in Eq.~\eqref{eq:supercharge_structure}~\footnote{We could have also chosen $\Gamma_j=\diag \{U_\mu \tau_j U_\mu^\dagger\}$ with $U_\mu=\exp{(i\alpha_\mu \tau_3)}$ having an arbitrary sequence of $\alpha_\mu$.
This does not mean that one can \emph{simultaneously} choose a continuum different supercharges: Mutual anticommutation allows two supercharges from this continuum, unitary equivalent to the choice $\Gamma_{1,2}=\tau_{1,2}$.}.

To study classes BDI and CI, we focus on a degeneracy space with energy $\varepsilon_\mu$ and first establish the form of $T_\pm$ and thus $Z$ in this space.
$T_+^2=+1$ implies that parity is again the only degeneracy, so $T_- = \tau_{1(2)} \mathcal{K}$ in class BDI (CI). 
$T_+\ket{\psi_\mu^p} \propto \ket{\psi_\mu^p}$ implies that the most general form is $T_+=\exp(i\varphi_{\mu}\tau_{3})\mathcal{K}$. 
With a suitable choice of the relative phases between the two parity sectors we can thus use $Z=T_+ T_-=\tau_1$; in this basis $T_+=\mathcal{K}$ ($T_+=\tau_3\mathcal{K}$) for class BDI (CI). 
The two $\Gamma_j$ satisfying Eq.~\eqref{eq:Gamma} can again be chosen as $\Gamma_{1,2}=\tau_{1,2}$.
However, checking the (anti)commutation with $Z$ we find that only $\Gamma_1$ is local.
Conversely, we can identify $\Gamma_2\equiv\Gamma_\infty\equiv \gamma_\infty$; this is consistent both with $\gamma_\infty$ itself satisfying Eq.~\eqref{eq:Gamma} and its transformation under $T_+$. 
We thus find $\mathcal{N}_\text{loc}=1$.

In classes DIII and CII, we find $\mathcal{N}_\mathrm{loc} =3$ local supercharges as we show in detail in Appendix~\ref{sec:supercharges}.
The spectrally flattened supercharges can be written as Kronecker products $\Gamma_j =\tau_1 \sigma_j$.
Their product $\Gamma_4 = -i \Gamma_1 \Gamma_2 \Gamma_3 = \tau_1$ is also local, but does not anticommute with $\Gamma_{j \le \mathcal{N}_\mathrm{loc}}$; it does, however, contribute to correlation function, as we discuss in the following.
As in classes BDI and CI, the nonlocal supercharge is $\Gamma_\infty = \tau_2$.

The values $\mathcal{N}_\text{loc}$, together with the sign $s$ in $T_\pm \Gamma_{j\leq\mathcal{N}_\text{loc}}T_\pm^{-1} = s\Gamma_{j\leq\mathcal{N}_\text{loc}}$ have a natural interpretation if one views the SYK model as arising at the end of a one-dimensional topological phase in class BDI~\cite{Fidkowski:2011dh,You:2017jj}.
These systems admit a $\mathbb{Z}_8$ classification:
At one of their ends, they have $k_s$ Majoranas satisfying $T_\pm \gamma_q T_\pm^{-1}=s\gamma_q$;
the topological index is $\nu= (k_+-k_-) \mod 8$.
Thus, the eight topological classes can be labeled by $\nu = 0,1,2,3,4,-3,-2,-1$ with the integers counting the number and sign of unpaired Majoranas.
In the SUSY classes, we find the same pattern for $s\mathcal{N}_\text{loc}$ against $k \mod 8$ ($T_\pm \gamma_{q\neq \infty} T_\pm^{-1}=\gamma_{q \neq \infty}$ implies $k_+=k$, $k_-=0$), see Table~\ref{tab:supercharge_info}.
The $\mathcal{N}_\text{loc}$ supercharges $\Gamma_{j\leq\mathcal{N}_\text{loc}}$ can thus be viewed as the many-body emergence of the minimal number and type of unpaired Majoranas consistent with $k$.

Next we turn to the structure of the supercharges in terms of the Majorana fermions $\gamma_q$.
For this, we employ another operator basis of the Hilbert space, the products of $n_a$ Majorana operators $\gamma_{q\neq\infty}$~\cite{Goldstein:2012ci}
\begin{equation}
 \Upsilon_a = i^{n_a (n_a -1)/2} \gamma_{i_1 (a)} \gamma_{i_2 (a)} \cdots \gamma_{i_{n_a} (a)}
 \label{eq:upsilon}
\end{equation}
with $i_j (a) \neq i_{j'} (a)$.
$\Upsilon_a$ are Hermitian, unitary, and orthonormal with respect to the trace, $\tr \left[ \Upsilon_a \Upsilon_b \right] /M = \delta_{ab}$.
In total, there are $2^{k}$ local operators $\Upsilon_a$~\cite{Goldstein:2012ci}.
As we aim to expand $\Gamma_{j\neq \infty}$, i.e., Hermitian odd-parity operators in terms of $\Upsilon_a$, we use only those $\Upsilon_a$ with odd $n_a$, and use only real expansion coefficients. 

Both time-reversal and particle-hole symmetry have the same (anti-) commutation properties when acting on $\Upsilon_a$.
Since $T_\pm \gamma_q T_\pm^{-1} = \gamma_q$, only the phase of $\Upsilon_a$ [cf.\ Eq.~\eqref{eq:upsilon}] may change when applying $T_\pm$, giving
 \begin{equation}
  T_\pm \Upsilon_a T_\pm^{-1} = (-1)^{n_a (n_a -1)/2} \Upsilon_a.
  \label{eq:anti_commutation}
\end{equation}
That is, $T_\pm$ and $\Upsilon_a$ commute when $n_a = 4n + 1$, and anticommute when $n_a = 4n +3$.
This, together with $v_{j,a}\in \mathbb{R}$ below, implies that, when expanding the supercharges,
\begin{equation}
 \Gamma_j = \sum_a v_{j,a} \Upsilon_a ,\quad \sum_a v_{j,a}^2=1,
\label{eq:expansion_supercharge}
\end{equation}
only terms with $n_a = 4 n+1$ contribute to $\Gamma_j$ when $[T_\pm ,\Gamma_j] = 0$, and only terms with $n_a = 4n+3$ contribute when $\lbrace T_\pm,\Gamma_j \rbrace =0$.
In classes DIII and CII we also consider $\Gamma_4=-i\Gamma_1\Gamma_2\Gamma_3$ whose transformation properties follow from those of $\Gamma_{1,2,3}$.
The resulting expansion structure is summarized in Table.~\ref{tab:supercharge_info}. 

\begin{table}
 \begin{tabular}{c|cccccc}
 \toprule
  $k \mod 8 $ & 1 & 2 & 3 & 5 & 6 & 7 \\
  \colrule
  Label & BDI & D & DIII & CII & C & CI \\
  $\beta$ & $1$ & $2$ & $4$ & $4$ & $2$ & $1$ \\
  \colrule
  $s\mathcal{N}_\text{loc}$	& $1$ & $2$ & $3$ & $-3$ & $-2$ & $-1$ \\
  $\Gamma_{j\leq\mathcal{N}_\text{loc}}$	& $4n+1$		& $4n+1$ & $4n+1$ & $4n+3$ & $4n+3$ & $4n+3$	\\
  \colrule
  $\Gamma_{4}$	&	&	& $4n+3$ & $4n+1$ &	&	\\
  \botrule
 \end{tabular}
 \caption{The Dyson index $\beta$, number $\mathcal{N}_\text{loc}$,  signature $T_\pm \Gamma_{j\leq\mathcal{N}_\text{loc}}T_\pm^{-1}=s\Gamma_{j\leq\mathcal{N}_\text{loc}}$, and the Majorana fermion structure of $\Gamma_{j\neq \infty}$.
 (The supercharges $Q_j$ have the same properties, since \mbox{$T_\pm HT_\pm^{-1}=H$.})
 In the Majorana expansion of $\Gamma_j$, only those $\Upsilon_a$ with $n_a = 4n+1$ or $n_a = 4n+3$ contribute; the two options are shown in the last two rows of the table.
 The horizontal line visually distinguishes $\Gamma_4$ from the three supercharges because it does not anticommute with them.
 A blank entry indicates that $\Gamma_4$ does not exist in these classes.
 }
 \label{tab:supercharge_info}
\end{table}

Having discussed the interplay of SUSY and the symmetry classification, we now identify signatures of SUSY, $\mathcal{N}_\text{loc}$, and the supercharge structure in various observables.
A simple link between $\mathcal{N}_\text{loc}$ and observables exists due to the fact that the number of different $\Gamma_{j\leq \mathcal{N}_\text{loc}}$ and their linearly independent odd-parity products, i.e., including $\Gamma_4$ in classes CII and DIII, equals the degrees of freedom $\beta$ (i.e., the Dyson index linked to $T_+$~\cite{Haake:2010hc}) of the Hamiltonian's off-diagonal matrix elements.
In fact, the most general Hermitian linear combinations of these $\Gamma_j$ have the same type of offdiagonals, up to an imaginary unit, as the Hamiltonian:
real for $\beta=1$ (classes BDI and CI), complex for $\beta=2$ (classes D and C) and real quaternion for $\beta=4$ (classes DIII and CII).

\begin{figure}
 \includegraphics[scale=1]{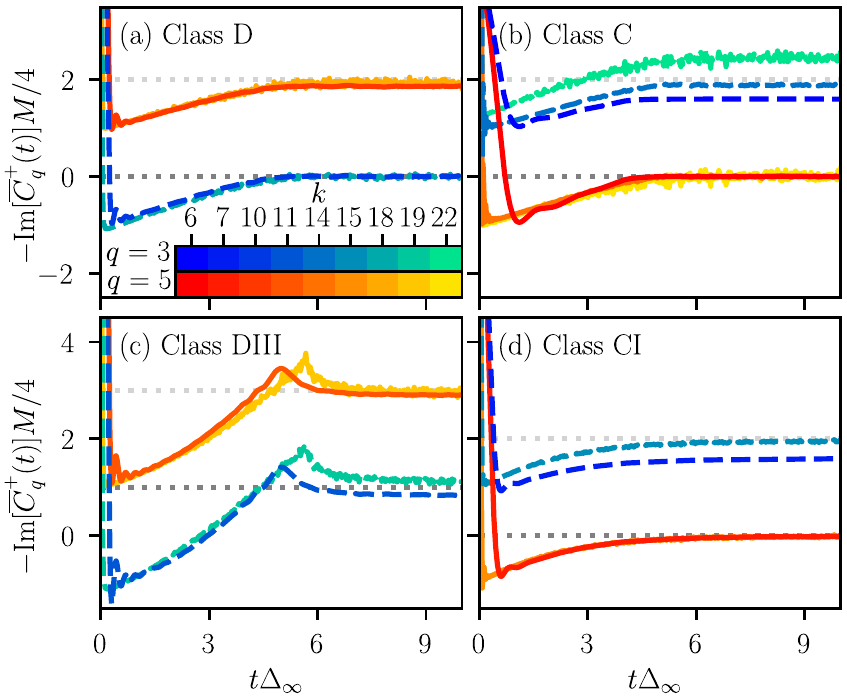}
 \caption{$q$-body time-dependent correlation function at infinite temperature, averaged over an ensemble of up to $2^{16}$ Gaussian distributed $J_{qrst}$, for classes (a) D, (b) C, (c) DIII, and (d) CI.
 The different colors denote different $k$ and $q$, cf.\ inset in~(a), the dashed lines represent $q=3$ and the solid lines $q=5$.
 The ramp shape follows the Dyson index $\beta$ and hence links to the number of supercharges.
 The long-time plateau $\overline{C}_{q,\infty}$ is studied in Fig.~\ref{fig:weight_qbody}.
 Error bars are either smaller than the line width (for small $k$) or smaller than the disorder-induced fluctuations of the lines (for large $k$).
 }
 \label{fig:time_evolution}
\end{figure}

In the SUSY classes, the value of $\beta$ sets the energy level correlations, including the long-range spectral rigidity, across \emph{opposite} parity sectors (these are uncorrelated without SUSY) which lead to ``ramps'' in time-dependent correlation functions of parity-\emph{odd} observables.
(For single-Majorana examples see Refs.~\onlinecite{Cotler:2017fx,Behrends:2019jc}.)
These ramps occur at time scales below $2\pi$ times the inverse mean level spacing $1/\Delta_\infty$, and have $\beta$-dependent shape~\cite{Guhr:1998bg}.
In particular, the ramp connects to a long-time plateau smoothly when $\beta=1$, sharply when $\beta=2$, and with a kink when $\beta=4$.
In Fig.~\ref{fig:time_evolution}, we show ensemble-averaged $q$-body correlation functions [Eq.~\eqref{eq:correlation_function} below] in classes D, C, DIII, and CI.
For completeness, we show the correlation function in the remaining symmetry classes, including those that do not support SUSY, in Appendix~\ref{sec:correlations}.

\begin{figure}
 \includegraphics[scale=1]{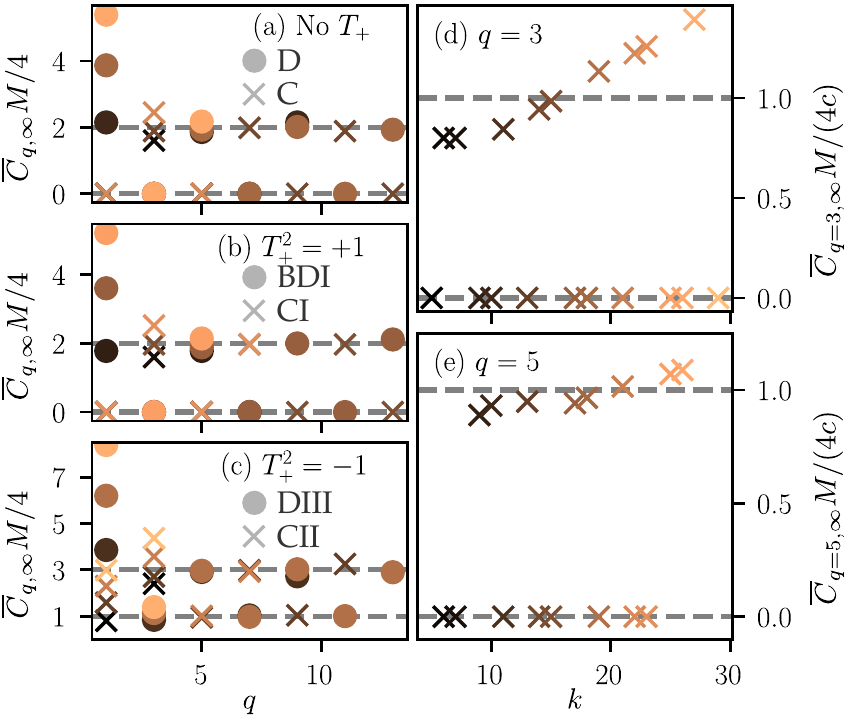}
 \caption{
 Normalized plateau $\overline{C}_{q,\infty} M/4$ of the $q$-body correlation function, averaged over an ensemble of up to $2^{14}$ Gaussian distributed $J_{qrst}$.
 The color encodes the number $k$ of Majoranas, cf.\ panels~(d) and~(e).
 In all classes, $\overline{C}_\infty M/4$ alternates with $q$ approximately as predicted in Eq.~\eqref{eq:CinftyRMT}; the agreement is excellent when $\binom{k}{q}/\binom{k}{\lfloor k/2 \rfloor}$ is close to one.
 In panel~(d), we show that $\overline{C}_\infty M/ (4c)$ [with $c$ the random matrix expectation based on Eq.~\eqref{eq:CinftyRMT}] increases as a function of $k$, but with a rate that decreases upon increasing $q$ [panel (e)].
 Statistical error bars are smaller than the marker size.
 }
 \label{fig:weight_qbody}
\end{figure}

Besides this direct correspondence between the supercharges and ramp structure, we additionally find more subtle consequences of SUSY:
The long-time ($t \gg 1/\Delta_\infty$) plateau in $q$-body correlation functions is also related to the number and structure of the supercharges, cf.\ Fig.~\ref{fig:weight_qbody}.
To quantify this relationship, we consider the retarded time-dependent $q$-body correlation function
\begin{equation}
 C_{q}^+ (t) = - i \Theta (t) \frac{1}{\binom{k}{q}} \sum_{a,n_a =q} \langle \lbrace \Upsilon_a (t) , \Upsilon_a (0) \rbrace \rangle ,
\end{equation}
where $\langle \ldots \rangle$ denotes thermal average.
Although the signatures we reveal are present at any temperature, we find an especially transparent relationship at infinite temperature, where the correlation function reads
\begin{align}
 C_{q}^+ (t) =& -i \Theta (t) \frac{1}{\binom{k}{q}} \sum_{a,n_a =q} \frac{1}{M} \sum_{p\mu\nu} \left| \bra{\psi_\mu^p} \Upsilon_a \ket{\psi_\nu^{-p}} \right|^2 \nonumber \\
 & \times 2 \cos \left( t \left( \varepsilon_\mu^p - \varepsilon_\nu^{-p} \right) \right).
\label{eq:correlation_function}
\end{align}
When $t \gg 1/\Delta_\infty$, terms with $\varepsilon_\mu^p \neq \varepsilon_\nu^{-p}$ give a quickly oscillating contribution $\delta C_q^+ (t)$ that averages to zero.
Only states with $\varepsilon_\mu^p = \varepsilon_\nu^{-p}$ give a time-independent contribution $C_{q,\infty}$.
Thus, $C_q^+ (t) = -i \Theta(t) [ C_{q,\infty} + \delta C_q^+ (t) ]$ with
\begin{equation}
 \!\!\!\! C_{q,\infty}=\frac{1}{\binom{k}{q}}\sum_{a,n_{a}=q}\frac{2}{M}\sum_{\mu} \tr \Upsilon_{a\mu}^{2},\quad\Upsilon_{a\mu}=P_{\mu}\Upsilon_{a}P_{\mu},
 \label{eq:weight}
\end{equation}
where in converting the equal-energy sum to a trace, we introduced the projection $P_{\mu}$ to the eigenspace with energy $\varepsilon_{\mu}$ and used that $\Upsilon_{a}$ is Hermitian and parity odd.

Next we convert Eq.~\eqref{eq:weight} into a sum over $\Gamma_{j<\infty}$.
We start by expanding $\Upsilon_{a\mu}=\sum_{j<\infty}y_{j}P_{\mu}\Gamma_{j}$ (with real $y_j$) which holds as within an eigenspace, the (projected) operators $P_{\mu}\Gamma_{j<\infty}$ form a basis for local, Hermitian, parity-odd operators.
If $\Upsilon_{a}$ transforms the same (opposite) way to $\Gamma_{j}$ under $T_{\pm}$ then generically $y_{j}\neq0$ ($y_{j}=0$).
Now using the trace-orthogonality of the $\Gamma_{j<\infty}$ and $\tr \Gamma_{j}^{2} = 2^{\mathcal{N}/2}=\mathcal{N}$~(for $\mathcal{N}=2,4$), we find
\begin{equation}\label{eq:Gamma_exansion}
 C_{q,\infty}=\frac{1}{\binom{k}{q}} \sum_{a,n_{a}=q} \frac{2}{M\mathcal{N}} \sum_{\mu} \sum_{j<\infty} \left[ \tr( P_{\mu}\Upsilon_{a}P_{\mu}\Gamma_{j} ) \right]^{2}.
\end{equation}

A simple estimate for $C_{q,\infty}$ can be given assuming that expanding $P_\mu\Gamma_{j<\infty}=\sum_{a} v_{\mu j,a}\Upsilon_{a}$ results in random coefficients $v_{\mu j,a}$ subject only to normalization and symmetry constraints.
Denoting such a random vector average by $\overline{(\ldots)}$, we find
\begin{equation}\label{eq:CinftyRMT}
 \frac{\overline{C}_{q,\infty} M}{4}=\begin{cases}
 \frac{\mathcal{N}}{\beta}\mathcal{N}_{\text{loc}} & q:\,\Upsilon_{a}\triangleq\Gamma_{j\leq\mathcal{N}_{\text{loc}}},\\
 \frac{\mathcal{N}}{\beta}\delta_{\beta,4} & \text{otherwise},
\end{cases}
\end{equation}
where $\Upsilon_{a} \triangleq \Gamma_j$ here means that $\Upsilon_{a}$ transforms the same way as $\Gamma_j$ under $T_\pm$.
Thus, each $\Gamma_{j<\infty}$ give the same contribution to $\overline{C}_{q,\infty}$ when they contain $q$-Majorana terms and zero otherwise.
The nonzero value for $\beta=4$ when $\Upsilon_{a}\not\triangleq \Gamma_{j\leq \mathcal{N}_\text{loc}}$ arises due to $\Gamma_4$ since $\Gamma_4\not\triangleq\Gamma_{j\leq \mathcal{N}_\text{loc}}$.
Considering the Majorana structure of $\Gamma_j$ in Table~\ref{tab:supercharge_info}, Eq.~\eqref{eq:CinftyRMT} translates to an alternating pattern of $\overline{C}_{q,\infty}$ as $q$ is varied in a given symmetry class, with complementary $\overline{C}_{q,\infty}$ values for classes with opposite $s\mathcal{N}_\text{loc}$.

In Fig.~\ref{fig:weight_qbody}, we show the numerically obtained value of $\overline{C}_\infty M/4$ for various $k$ and $q$.
The alternating pattern expected from Eq.~\eqref{eq:CinftyRMT} is clearly visible [panels~(a) to~(c)].
While the numerical value of the nonzero plateau differs from expectation when $q \ll k$ (and $k-q \ll k$, not shown), Eq.~\eqref{eq:CinftyRMT} gives an accurate prediction when $\binom{k}{q}/\binom{k}{\lfloor k/2 \rfloor}$ is close to one (with $\lfloor \ldots \rfloor$ the floor function).
To investigate this further, in panels~(d) and~(e), we show $C_\infty M/4$ versus $k$.
The growth with $k$ is slower for $q=5$ than for $q=3$, which is in turn slower than the almost linearly growing $q=1$ case~\cite{Behrends:2019jc}.

To summarize, we have shown that supersymmetry is (almost) always present in the SYK model with generic four-body interactions.
It is only absent in those classes without particle-hole symmetry, i.e., in classes AI and AII.
The type of SUSY, in particular the number $\mathcal{N}_\text{loc}$ of local supercharges and their symmetry properties follow a pattern that finds a natural interpretation when $\Gamma_{j\le \mathcal{N}_\mathrm{loc}}$ are viewed as emergent Majorana fermions in a one-dimensional topological phase with SYK model boundary physics. 
These SUSY features all link directly to features in time-dependent correlation functions of fermion-parity-odd observables. 
For $q$-body retarded correlation functions, this includes the shape of the ramp in the short-time regime, due to a link between $\mathcal{N}_\text{loc}$ and the Dyson index $\beta$; and the value of the long-time plateau due to the imprint of how $\Gamma_j$ transforms under $T_\pm$ on its microscopic Majorana structure. 
These $q$-body correlation functions, even with large $q$, can be measured in digital quantum simulation of the SYK model~\cite{Garcia:2017bb}.
The single-particle Green's function ($q=1$) is accessible through scanning tunneling spectroscopy~\cite{Pikulin:2017js,Chew:2017fn}.
We stress that the features in the correlation functions are dynamical consequences of SUSY, which are less frequently considered than ground-state consequences~\cite{Fendley:2011iv,Cubero:2016dm}.

\begin{acknowledgments}
We thank David~Tong for helpful discussions.
This work was supported by the ERC Starting Grant No.\ 678795 TopInSy.
\end{acknowledgments}

\appendix

\section{Eightfold symmetry classification of the SYK model}
\label{sec:eightfold}

In the main text, we make extensive use of the eightfold symmetry classification of the SYK model with four-body interactions.
This classification, introduced in Ref.~\onlinecite{Behrends:2019jc}, reveals an Altland-Zirnbauer structure behind the period-eight pattern of Wigner-Dyson classes highlighted in Ref.~\onlinecite{You:2017jj}.
The considerations of both Refs.~\onlinecite{Behrends:2019jc,You:2017jj} are based on the work of Fidkowski and Kitaev on the topological classification of interacting fermions in one dimension~\cite{Fidkowski:2011dh}, which Ref.~\onlinecite{You:2017jj} noted to apply to the SYK model upon viewing it as existing at the end of a one-dimensional fermionic topological phase.
Here we summarize the key ideas in Ref.~\onlinecite{Behrends:2019jc} and highlight certain aspects relevant for the main body of the paper.

The eightfold symmetry classification is based on the presence of antiunitary symmetries and their relation to fermion parity.
First note that the SYK Hamiltonian [Eq.~(2) in the main text] conserves fermion parity, $[H,P]=0$ with
\begin{equation}
 P =
 \begin{cases}
  i^{k/2} \gamma_1 \gamma_2 \cdots \gamma_k & \text{ even $k$} \\
  i^{(k+1)/2} \gamma_1 \gamma_2 \cdots \gamma_k \gamma_\infty & \text{ odd $k$}
 \end{cases}
 \label{eq:parity}
\end{equation}
where the Majorana $\gamma_\infty$ ensures a well-defined fermion parity for odd $k$, but does not contribute to local operators, including the Hamiltonian.
All Majorana operators including $\gamma_\infty$ are odd in fermion parity and thus satisfy $\{P,\gamma_q \} =0$.
When $k$ is odd, the Hamiltonian also commutes with the operator
\begin{equation}
 Z = - i^{(k-1)/2} \gamma_1 \gamma_2 \cdots \gamma_k
\end{equation}
that in turn commutes with all local Majorana operators $[Z,\gamma_{q \neq \infty}]=0$, but anticommutes with the additional Majorana at infinity $\{ Z,\gamma_{\infty} \}=0$ and accordingly with fermion parity $\{ Z, P\} = 0$.

Depending on the number of Majorana operators, it may be possible to find local unitary operators $C_\pm$ that satisfy~\cite{de1986field}
\begin{equation}
 C_\pm \gamma_{q \neq \infty}^* C_\pm^\dagger = \pm \gamma_{q \neq \infty} .
\end{equation}
In other words, the antiunitary operator $T= C_+ \mathcal{K}$ commutes with all local Majorana operators, whereas $A = C_- \mathcal{K}$ anticommutes with them.

For even $k$, it is always possible to find both operators $C_+$ and $C_-$~\cite{de1986field}.
Using Eq.~\eqref{eq:parity} we conclude
\begin{equation}
 T P T^{-1} = (-1)^{k/2} P,
\end{equation}
i.e., $T$ commutes with fermion parity when $k\mod 4 = 0$ and anticommutes with fermion parity when $k\mod 4 = 2$.
Thus, when both operators commute, $T$ is an antiunitary operator that maps each parity subblock of the Hamiltonian to itself, otherwise, the parity subblocks are exchanged by $T$.
We refer to the first case as time-reversal symmetry: The level spacing statistics of each subblock are determined by the presence of $T$ and by the sign $T^2 = \pm 1$.
We refer to the second case as particle-hole symmetry, as the parity sectors are exchanged.
Particle-hole symmetry sets correlations {\it across} different parity sectors, a feature we exploit in the main text.
In the following, we distinguish the two cases by the notation $T_+$ for time-reversal ($[T_+,P]=0$) and $T_-$ for particle-hole symmetry ($\{ T_-,P \} =0$), the same notation we use in the main text.

For odd $k$, only one of two operators $C_\pm$ is local.
In particular, when $k =4n+1$, only $C_+$ is local, and when $k = 4n+3$, only $C_-$ is local (with integer $n$).
When only $C_-$ is local, we can use fermion parity to define $T= P C_- \mathcal{K}$, an antiunitary operator that commutes with all fermions; when $C_+$ is local, we can use $T=C_+ \mathcal{K}$.
This implies that $T$ and $P$ commute
\begin{align}
 T P T^{-1}
 &= (-1)^{(k+1)/2} \gamma_1 \gamma_2 \cdots \gamma_k T \gamma_\infty T^{-1} \\
 &= P,
\end{align}
where we chose $T \gamma_\infty T^{-1} = (-1)^{(k+1)/2} \gamma_\infty$ for convenience.
We thus identify $T = T_+$.
Due to the presence of $Z$, we can define a second antiunitary operator $T_-$ via $T_- = T_+^{-1} Z$ such that $Z = T_+ T_-$ as quoted in the main text. Since $Z$ anticommutes with fermion parity, $T_-$ also anticommutes with it, $\{ T_- ,P \} =0$.

The squares of $T_+$ and $T_-$ can be determined from their explicit construction in terms of Majorana operators~\cite{Fidkowski:2011dh}.
Without loss of generality, we choose the Majorana operators such that $\gamma_{2n+1}^* = \gamma_{2n+1}$ are real while $\gamma_{2n}^* = -\gamma_{2n}$ are purely imaginary.

When $k$ is even, the basis choice introduced above results in
\begin{equation}
 C_+ =
 \begin{cases}
  \gamma_2 \gamma_4 \cdots \gamma_{k}   & ~ k = 4 n \\
  \gamma_1 \gamma_3 \cdots \gamma_{k-1} & ~ k = 4 n +2 ,
 \end{cases}
\end{equation}
thus, $C_+$ is always the product of $k/2$ Majorana operators.
In both cases, $C_+ = C_+^*$ is real ($C_+$ is either the product of only real Majorana operators or the product of an even number of purely imaginary operators).
When $k=4n$, we have
\begin{align}
 T_+^2 = C_+^ 2 = \gamma_2\gamma_4 \cdots \gamma_k \gamma_2 \gamma_4 \cdots \gamma_k ,
\end{align}
giving $ T_+^2 = (-1)^{k/4}$, and when $k=4n+2$,
\begin{align}
 T_+^2 = C_+^ 2 = \gamma_1\gamma_3 \cdots \gamma_{k-1} \gamma_1 \gamma_3 \cdots \gamma_{k-1} ,
\end{align}
which gives $ T_+^2 = (-1)^{(k-2)/4}$.

When $k$ is odd, we need to distinguish the two cases $k=4n+1$ and $k=4n+3$.
For $k=4n+1$ we find
\begin{equation}
 C_+ = \gamma_2 \gamma_4 \cdots \gamma_{k-1} .
\end{equation}
Since $k$ is odd, the nonlocal Majorana $\gamma_\infty^* = -\gamma_\infty$ is purely imaginary, giving
\begin{equation}
 C_+ \gamma_\infty^* C_+^\dagger = -(-1)^{(k-1)/2} \gamma_\infty  = (-1)^{(k+1)/2} \gamma_\infty
\end{equation}
where we used that $C_+$ contains $(k-1)/2$ Majorana operators $\gamma_{q \neq \infty}$.
Accordingly, we identify $T_+ = C_+ \mathcal{K}$ and conclude $T_+^2 = (-1)^{(k-1)/4}$.
We further identify $T_- = T_+^{-1} Z$ with $T_-^2 = T_+^2 T_+ Z T_+^{-1} Z$.
Using $T_+ Z T^{-1} = (-1)^{(k-1)/2} Z = Z$ and $Z^2 = 1$, we realize that $T_-^2 = T_+^2$.

For $k = 4n+3$ we find
\begin{equation}
 C_- = \gamma_1 \gamma_3 \cdots \gamma_{k} ,
\end{equation}
which is a product of $(k+1)/2$ Majorana operators. Since $C_- \mathcal{K}$ and $P$ both anticommute with all local Majorana operators, the antiunitary operator $P C_- \mathcal{K}$ commutes with them.
Again, $\gamma_\infty^* = - \gamma_\infty$, which gives
\begin{equation}
 P C_- \mathcal{K} \gamma_\infty (P C_- \mathcal{K})^{-1} = (-1)^{(k+1)/2} \gamma_\infty,
\end{equation}
where we used that $C_-$ is the product of $(k+1)/2$ local Majorana operators and $\{ P,\gamma_\infty \} = 0$.
Thus, we identify $T_+ = P C_- \mathcal{K}$.
The square $T_+^2 = P C_- P^* C_- = (-1)^{k+1} C_-^2$ and $C_-^2 = (-1)^{(k+1)/4}$, such that $T_+^2 = (-1)^{(k+1)/4}$.
Accordingly, $T_-^2 = T_+^2 T_+ Z T_+^{-1} Z = (-1)^{(k-1)/2} T_+^2 = - T_+^2$.

We summarize these results in Table~I in the main text.

\section{Supercharges in classes DIII and CII}
\label{sec:supercharges}

In this section, we construct the supercharges in classes DIII and CII along the lines of the same strategy that we followed in the main text for the other classes.
We show and discuss how the supercharge structure summarized in the main text arises in these classes.

In classes DIII and CII, time reversal symmetry with $T_+^2 = -1$ implies Kramers degeneracy:
The states $\ket{\psi_\mu^p}$ and $T_+ \ket{\psi_\mu^p}$ are orthogonal.
Together with parity degeneracy this gives a fourfold degeneracy of energy eigenvalues and $\mathcal{N}=4$.
In its diagonalized form, the Hamiltonian reads $H=\diag (\{ \varepsilon_\mu \} ) \otimes \openone_{2^{\mathcal{N}/2}}$ with $\openone_{2^{\mathcal{N}/2}}=\tau_0\sigma_0$, where the matrices $\tau_\mu$ act in parity grading space and $\sigma_\nu$ in the space of Kramers doublets (hence $\tau_\mu \sigma_\nu$ is a shorthand for the Kronecker product $\tau_\mu \otimes \sigma_\nu$; additional tensor products with the identity are implied).

Without loss of generality, we can choose $Z=\tau_1$. 
The four $\Gamma_j$ satisfying Eq.~(5) in the main text are now the familiar Dirac matrices, chosen as $\Gamma_j=\tau_1\sigma_j$ ($j=1,2,3$) and $\Gamma_\infty=\tau_2$.
Checking the (anti)commutation with $Z$ we indeed find that only $\Gamma_{1,2,3}$ are local.
Conversely, identifying $\Gamma_\infty\equiv \gamma_\infty$ and requiring that it transforms under $T_+$ as $T_+ \gamma_\infty T_+^{-1} = (-1)^{(k+1)/2} \gamma_\infty$ sets
\begin{equation}
T_+= \begin{cases}
i\tau_3\sigma_2\mathcal{K} & \text{ class DIII} \\
i\sigma_2\mathcal{K} & \text{ CII}
\end{cases}
\end{equation}
in this basis.
The form of particle-hole symmetry $T_-$ follows from $T_+ T_-= Z$.

The product $\Gamma_4 = -i \Gamma_1 \Gamma_2 \Gamma_3$ is also Hermitian, parity odd, local, and linearly independent of $\Gamma_{1,2,3}$.
Furthermore, $Q_4= \diag ( \lbrace \sqrt{\varepsilon_\mu} \rbrace ) \otimes \Gamma_4$ squares to the Hamiltonian.
It is, however, not a supercharge because $\Gamma_4$ does not anticommute with $\Gamma_{1,2,3}$.
(Nevertheless, $\Gamma_4$ contributes to correlation functions, as discussed in the main text.)
We thus find $\mathcal{N}_\text{loc}=3$.

Using the explicit form of $T_+$ obtained above and $T_- = T_+^{-1} Z$, we find
\begin{equation}
 T_\pm \Gamma_{j \le \mathcal{N}_\mathrm{loc}} T_\pm^{-1}  =
 \begin{cases}
  +\Gamma_{j \le \mathcal{N}_\mathrm{loc}} & \text{ class DIII} \\
  -\Gamma_{j \le \mathcal{N}_\mathrm{loc}} & \text{ class CII.}
 \end{cases}
\end{equation}
This implies that
\begin{equation}
 T_\pm \Gamma_4 T_\pm^{-1} = - \Gamma_4
 = \begin{cases}
 - \Gamma_4 & \text{ class DIII} \\
 + \Gamma_4 & \text{ class CII.}
 \end{cases}
\end{equation}

\begin{figure}
\includegraphics{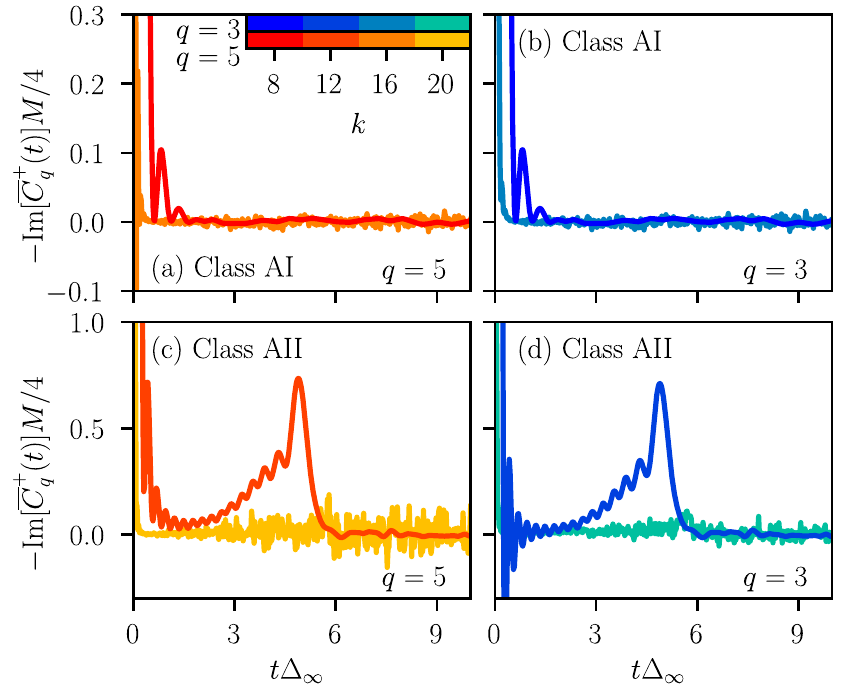}
\caption{Ensemble-averaged $q$-body correlation function in symmetry classes AI [$k \mod 8 =0$, panels~(a) and~(b)] and AII [$k \mod 8 =4$, panels~(c) and~(d)].
The color code denotes the different values of $k$ and $q$; cf.\ the inset in panel~(a).
Here, $q=5$ serves is a representative example for $q=4n+1$, and $q=3$ for $q=4n+3$.
The correlation function decays to zero in all cases shown here. Only when $k=12$ [panels~(c) and~(d)] do the two parity sectors show correlations, which is a small-size effect.
All results are averaged over a large ensemble, ranging from $768$ (for $k=20$, $q=5$) to $2^{17}$ (for $k=8$) realizations of the couplings $J_{qrst}$.
Error bars are either smaller than the line width (for small $k$) or smaller than the disorder-induced fluctuations of the lines (for large $k$).
}
\label{fig:time_evolution_AI_AII}
\end{figure}

For completeness, we give these operators in terms of the eigenstates. To give explicit relations, we need to fix certain phase relations between the eigenstates. Defining time-reversal symmetry as
\begin{equation}
 T_+ \ket{\psi_{2\mu+1}^+} = \ket{\psi_{2\mu}^+}
\end{equation}
fixes the phase relation between Kramers doublets that we denote by $\ket{\psi_{2\mu}^p}$ and $\ket{\psi_{2\mu+1}^p}$.
This choice sets $\varepsilon_{2\mu} = \varepsilon_{2\mu+1}$.
We fix the phase between opposite parities by choosing
\begin{equation}
 Z \ket{\psi_{\mu}^p} = \ket{\psi_{\mu}^{-p}} .
\end{equation}
In this basis, the supercharges defined above read
\begin{align}
 \Gamma_1 &= \sum_{p\mu} \left( \ket{\psi_{2\mu}^p} \bra{\psi_{2\mu+1}^{-p}} + \ket{\psi_{2\mu+1}^p} \bra{\psi_{2\mu}^{-p}} \right) \\
 \Gamma_2 &= -i \sum_{p\mu} \left( \ket{\psi_{2\mu}^p} \bra{\psi_{2\mu+1}^{-p}} - \ket{\psi_{2\mu+1}^p} \bra{\psi_{2\mu}^{-p}} \right) \\
 \Gamma_3 &= \sum_{p\mu} \left( \ket{\psi_{2\mu}^p} \bra{\psi_{2\mu}^{-p}} - \ket{\psi_{2\mu+1}^p} \bra{\psi_{2\mu+1}^{-p}} \right) .
\end{align}
Their product is accordingly
\begin{align}
 \Gamma_4
 &= -i \Gamma_1 \Gamma_2 \Gamma_3
 = \sum_{p\mu} \ket{\psi_{\mu}^p} \bra{\psi_{\mu}^{-p}} = Z
\end{align}
and the nonlocal supercharge
\begin{align}
 \Gamma_\infty &= -i \sum_\mu  \left( \ket{\psi_{\mu}^+} \bra{\psi_{\mu}^{-}} - \ket{\psi_{\mu}^-} \bra{\psi_{\mu}^{+}} \right) .
\end{align}
Writing the fermion parity $P$ in terms of the eigenstates
\begin{align}
 P &= \sum_\mu  \left( \ket{\psi_{\mu}^+} \bra{\psi_{\mu}^{+}} - \ket{\psi_{\mu}^-} \bra{\psi_{\mu}^{-}} \right)
\end{align}
and using $P=-i Z \gamma_\infty$ confirms that $\Gamma_\infty \equiv \gamma_\infty$.

\begin{figure}
\includegraphics{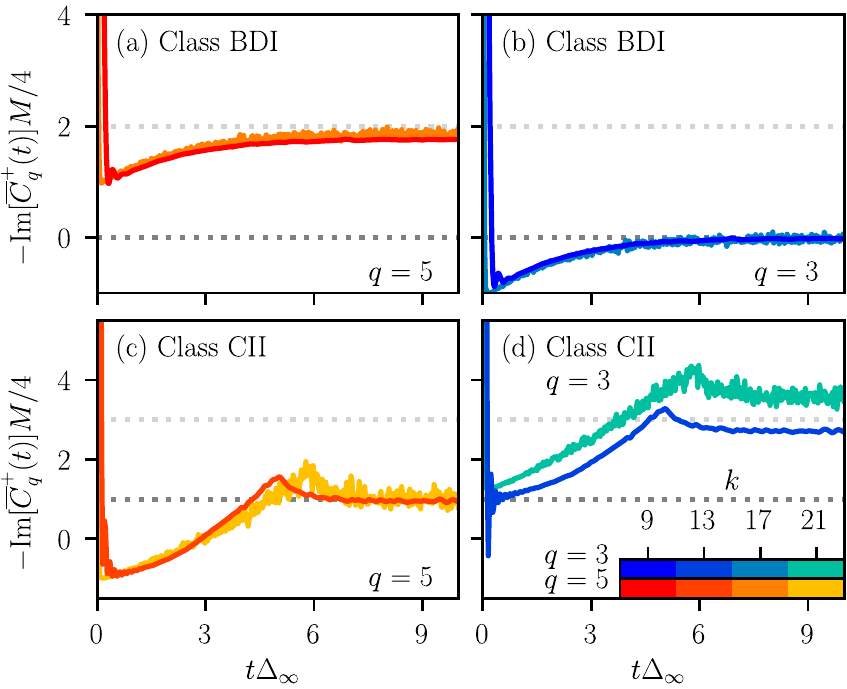}
\caption{Ensemble-averaged $q$-body correlation function in symmetry classes BDI ($k\mod 8=1$) and CII ($k\mod 8 = 5$).
In panels~(a) and~(b), we show different system sizes in class BDI, where the different colors denote $k$ and $q$; cf.\ inset in panel~(d).
As in Fig.~\ref{fig:time_evolution_AI_AII}, $q=5$ serves is a representative example for $q=4n+1$, and $q=3$ for $q=4n+3$.
In panels~(c) and~(d), we show different systems sizes in class CII.
The dotted lines show the expectation for $\overline{C}_{q,\infty}$ based on random matrix theory; cf.\ Eq.~(13) in the main text.
All results are averaged over a large ensemble, ranging from $192$ (for $k=21$, $q=5$) to $2^{14}$ (for $k=9$) realizations of the couplings $J_{qrst}$.
Similarly to Fig.~\ref{fig:time_evolution_AI_AII}, error bars are either smaller than the line width (for small $k$) or smaller than the disorder-induced fluctuations of the lines (for large $k$).
}
\label{fig:time_evolution_BDI_CII}
\end{figure}

\section{Correlation functions}
\label{sec:correlations}

\paragraph{Symmetry classes AI and AII.}
Classes AI ($k \mod 8 = 0$) and AII ($k \mod 8 =4$) are the two classes without SUSY.
In these classes, parity sectors are uncorrelated, hence, upon evaluating Eq.~(9) of the main text, we expect to see neither a ramp structure nor a plateau at long times $t \gg 1/\Delta_\infty$.
To support this expectation with numerical evidence, we show the ensemble-averaged $q$-body correlation function at infinite temperature [Eq.~(10) in the main text] in Fig.~\ref{fig:time_evolution_AI_AII}.
In Fig.~\ref{fig:time_evolution_AI_AII}(a), we show the $(q=5)$-body correlation function for $k=8,16$ Majoranas, i.e., symmetry class AI.
The correlation function rapidly decays to zero and does not exhibit a plateau. The correlation function behaves qualitatively similar when $q=3$, as shown in panel~(b).
The plateau $\overline{C}_{q,\infty} = 0$ is consistent Eq.~(13) in the main text since $\mathcal{N}=0$.

In Fig.~\ref{fig:time_evolution_AI_AII}(c) and~(d), we show the correlation function for $k=12,20$ Majoranas, i.e., symmetry class AII.
For $k=12$, we observe for both $q=3$ and $q=5$ that signatures appear at intermediate time scales ($t \approx \pi/\Delta_\infty$) before the correlation function eventually vanishes for longer times ($t \gg 1/\Delta_\infty$).
We attribute these correlations between the two parity sectors at intermediate time scales to small-size effects:
They are the time-domain signatures of spectral oscillations observed in Ref.~\onlinecite{Behrends:2019jc} and are present only for $k=12$, the smallest nontrivial instance of class AII, and already disappear for $k\ge 20$.
This finite-size effect is pronounced in class AII due to the strong level repulsion corresponding to the Dyson index $\beta=4$, in contrast to class AI with $\beta=1$.

\paragraph{Symmetry classes BDI and CII.} In Fig.~\ref{fig:time_evolution_BDI_CII}, we show the ensemble-averaged correlation functions in classes BDI and CII.
These classes are supersymmetric, therefore the correlation functions show the ramp and plateau structure discussed in the main text.
Indeed, we find a ramp that connects smoothly to a plateau in class BDI [Fig.~\ref{fig:time_evolution_BDI_CII}(a)--(b)].
When $q=4n+1$, this plateau is at $\overline{C}_{q,\infty} \approx 2$, otherwise, $\overline{C}_{q,\infty} =0$ and the correlation vanishes for long times $t\gg 1/\Delta_\infty$, as predicted by Eq.~(13) in the main text.
The smooth connection between ramp and plateau is characteristic for $T_+^2 = +1$~\cite{Guhr:1998bg}.
These features match the expectations for a single local supercharge (Dyson index $\beta=1$) that transforms under time-reversal and particle-hole symmetry as $T_\pm Q_1 T_\pm^{-1} = + Q_1$.

Similarly, we find a ramp that connects with a kink to a plateau in class CII [Fig.~\ref{fig:time_evolution_BDI_CII}(c)--(d)].
When $q=4n+1$, this plateau is at $\overline{C}_{q,\infty} \approx 1$, and when $q=4n+3$, it is at $\overline{C}_{q,\infty} \approx 3$, as predicted by Eq.~(13) in the main text.
The kink is characteristic for $T_+^2=-1$~\cite{Guhr:1998bg}.

The numerical data presented in this appendix thus supports the statements made in the main text, in particular, the direct relation between the number of supercharges and the Dyson index (that is reflected in the shape of the ramp), and the value of the plateau (that reflects the number of supercharges and their linearly independent products that transform in the same way under time-reversal and particle-hole symmetry as $\Upsilon_a$).

\bibliography{susy}

\end{document}